\documentclass{PoS}

\title{VHEeP: A very high energy electron--proton collider based on proton-driven plasma wakefield acceleration}

\ShortTitle{VHEeP: A very high energy electron--proton collider}

\author{A. Caldwell\\
        Max Planck Institute for Physics, Munich, Germany\\
        E-mail: \email{caldwell@mpp.mpg.de}}

\author{\speaker{M. Wing}\thanks{Also at Universit\"{a}t Hamburg and supported by DESY and the Alexander von 
Humboldt Foundation.}\\
        UCL, London, UK\\
        E-mail: \email{m.wing@ucl.ac.uk}}

\abstract{
Based on current CERN infrastructure, an electron--proton collider is proposed at a centre-of-mass energy of about 9\,TeV.   
A 7\,TeV LHC bunch is used as the proton driver to create a plasma wakefield which then accelerates electrons to 3\,TeV, 
these then colliding with the other 7\,TeV LHC proton beam. The basic parameters of the collider are presented, which although 
of very high energy, has integrated luminosities of the order of 1\,pb$^{-1}$/year. For such a collider, with a centre-of-mass 
energy 30 times greater than HERA, parton momentum fractions, $x$, down to about $10^{-8}$ are accessible for $Q^2$ of 
1\,GeV$^2$ and could lead to effects of saturation or some other breakdown of DGLAP being observed.  The total photon--proton 
cross section can be measured up to very high energies and also at different energies as the possibility of varying the electron 
beam energy is assumed; this could have synergy with cosmic-ray physics.  Other physics which can be pursued at such a 
collider are contact interaction searches, such as quark and electron substructure, and measurements of the proton structure 
as well as other more conventional measurements of QCD at high energies and in a new kinematic regime.  The events at very 
low $x$ will lead to electrons and the hadronic final state produced at very low angles and so a novel spectrometer device will be 
needed to measure these.  First ideas of the physics programme of such a collider are given.
}

\FullConference{XXIII International Workshop on Deep-Inelastic Scattering\\
		27 April - May 1 2015\\
		Dallas, Texas}

\begin{document}

\section{Introduction}

The HERA $eP$ machine is so far the only lepton--hadron collider worldwide.  Due to its centre-of-mass energy of about 300\,GeV, 
HERA dramatically extended the kinematic reach for the deep inelastic scattering process compared to previous fixed-target 
experiments~\cite{1506.06042}.  A broad range of physics processes and new insights were gleaned from HERA which complemented 
the $p\bar{p}$ and $e^+e^-$  colliders, the Tevatron and LEP.  
The LHeC project~\cite{lhec} is a proposed $eP$ collider with significantly higher energy and luminosity than HERA with a programme 
to investigate Higgs physics and QCD, to search for new physics, etc..  This will use significant parts of the LHC infrastructure at 
CERN with different configurations, such as $eA$, also possible.  These proceedings consider the possibility of having a 
very high energy electron--proton collider (VHEeP) with an $eP$ centre-of-mass energy of about 9\,TeV, a factor of six higher than proposed 
for the LHeC and a factor of 30 higher than HERA.

The VHEeP machine would strongly rely on the use of the LHC beams and the technique of plasma wakefield acceleration to 
accelerate electrons to 3\,TeV over relatively short distances.  Given such an acceleration scheme, the luminosities will be relatively low, 
$\mathcal{O}({\rm pb}^{-1})$.  Given such an increase in centre-of-mass energy, the VHEeP collider will probe a new regime in deep 
inelastic scattering, particularly at low values of the probed fractional parton momentum, $x$.  These proceedings will look at first 
investigations of the physics potential of such a collider as well as first ideas of its technical implementation.

\section{The VHEeP accelerator and detector}

Given the limitation in the accelerating gradient for RF cavities of about 100\,MV/m, due to structural breakdown, a future collider with 
electron energies at the TeV scale 
requires an accelerator of lengths 10s of kilometres.  Using a novel technique, plasma wakefield acceleration~\cite{prl:43:267}, gradients 
up to about 100\,GV/m have been observed~\cite{pwa}, although there are many challenges to be overcome before this scheme can 
be used in a real accelerator.  Plasma wakefield acceleration overcomes the limitation on the accelerating gradient as the plasma is already 
ionised itself.  The phenomenon of plasma wakefield acceleration relies on a compact particle or laser beam to drive the process in plasma.  
In the case of a proton beam, the free plasma electrons are attracted to the incoming beam, overshoot and create regions of high positive 
charge density due to the ions which have remained static during this process.  The plasma electrons are attracted to these regions of high 
positive charge and so an oscillating motion of the electrons is set up in which a strong accelerating gradient is initiated in the direction 
of the proton beam.  Given the high energy of existing proton beams, they could accelerate a witness bunch of electrons injected into the 
wakefield to the TeV scale~\cite{pdpwa}.  Proton-driven plasma wakefield acceleration has so far not been demonstrated experimentally 
and a collaboration, AWAKE, has been formed to show this using the CERN SPS beam~\cite{awake}.  The aim is to start taking data in 2016  
and, in an initial 2-year running period, demonstrate $\mathcal{O}(\rm GeV)$ acceleration of electrons within 10\,m of plasma.

\begin{figure}
\begin{center}
\includegraphics[width=0.25\textwidth]{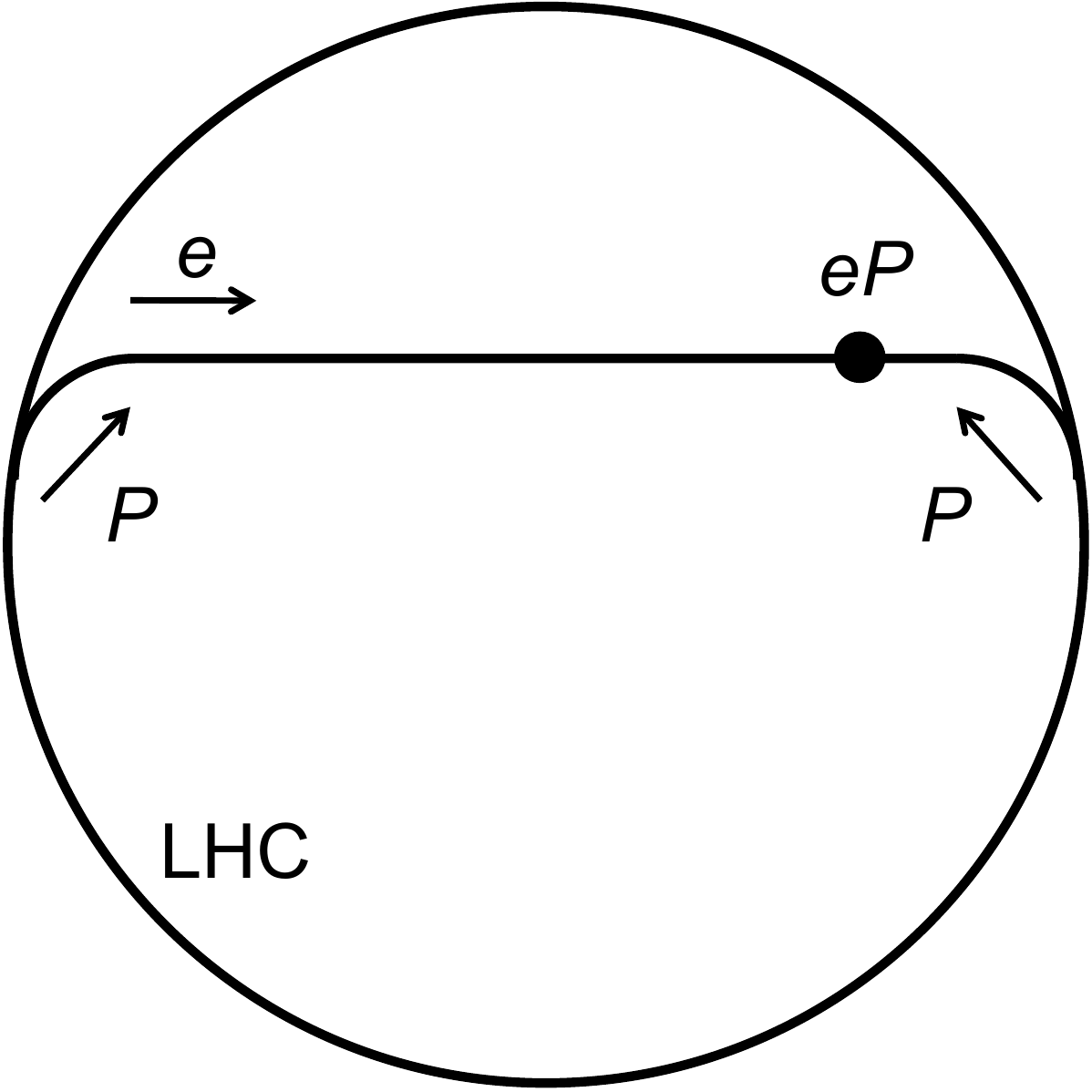}
\put(-110,100){(a)}
\hspace{1cm}
\includegraphics[width=0.65\textwidth]{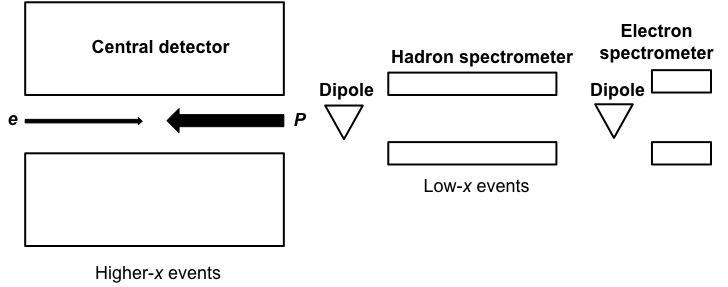}
\put(-290,100){(b)}
\end{center}
\caption{(a) Sketch of VHEeP collider.  One of the proton beams from the LHC is used as a driver to accelerator electrons 
which subsequently collide with the other LHC proton beam.  The $eP$ interaction point is indicated. (b) Sketch of VHEeP detector showing 
the need for a central, conventional collider detector and a long arm of spectrometer detectors to measure the scattered electrons 
and hadronic final state at low $x$.}
\label{fig:collidersketch}
\end{figure}

Simulations have shown that using the 7\,TeV LHC proton beam as designed, electrons can be 
accelerated to 3\,TeV in 4\,km~\cite{pp:18:103101}, about a factor of 10 less than needed for conventional RF acceleration.  For the 
VHEeP collider, such an 
electron beam is then assumed to collide with the other 7\,TeV LHC proton beam, giving an $eP$ centre-of-mass energy of about 9\,TeV.  
This is shown schematically in figure~\ref{fig:collidersketch}(a).  The emphasis is on using current CERN infrastructure where possible, i.e.\ 
the LHC with minimum modifications.  To obtain a new beamline within the LHC ring, high-gradient magnets will be needed to provide 
a smaller bending radius.  It is also assumed that the electron beam energy can be varied within the constraints of the length of the 
accelerator structure and so varying up to the maximum possible is assumed; this is motivated by the possibility to measure extra physics 
processes.

Although the energy of VHEeP will be very high, obtaining high luminosities will be, as with all plasma wakefield acceleration schemes, 
a challenge.  The luminosity is determined using the standard formula for a collider, 

$$
\mathcal{L} = \frac{f \cdot N_e \cdot N_P}{4 \, \pi \,\sigma_x \cdot \sigma_y}\,. 
$$
It is assumed that about 3000 bunches are delivered every 
30 minutes, giving a frequency, $f$, of 2\,Hz.  The number of protons, $N_P$, and number of electrons, $N_e$, per bunch are 
$4 \times 10^{11}$ and $1 \times 10^{11}$, respectively.  Assuming a beam cross section, $\sigma_x \sim \sigma_y$, of 4\,$\mu$m, 
gives a luminosity of $4 \times 10^{28}$\,cm$^{-2}$\,s$^{-1}$.  Running the collider for a large fraction of the year gives an integrated 
luminosity of about 1\,pb$^{-1}$/year.

Therefore, the expected luminosities are lower than were achieved at HERA and significantly lower than proposed for the LHeC.  
Hence, for the VHEeP facility, a physics case needs to be made for $eP$ physics at very high energies, but moderate luminosities and 
is the initial focus of the subsequent sections in these proceedings and will be for further studies of the VHEeP accelerator.

\section{Physics at VHEeP}

An initial list of some of the possible areas of investigation for a high energy $eP$ collider with moderate luminosities are:

\begin{itemize}

\item Measurements of cross sections at very low $x$ and investigation of saturation;

\item Investigation of contact interactions, e.g.\ radius of quarks and electrons;

\item Measurements of the total $\gamma P$ cross section at high energies and at many different energies.  This will allow a precise 
determination of its energy dependence and its relation to cosmic-ray physics can be investigated;

\item Investigation of the structure of the proton and photon, in particular the longitudinal proton structure function, $F_L$, which 
will also profit from the possibility to change the beam energy as well as results from $eA$ scattering;

\item Further tests of QCD such as extraction of the strong coupling at high energies.

\end{itemize}

In these proceedings, the first is investigated in more detail, whilst the others will be pursued in the future.  This list is not exhaustive 
and there may be other interesting possibilities.

In order to investigate the results of $eP$ collisions at a centre-of-mass energy of 9\,TeV, the {\sc Ariadne~4.12} Monte Carlo programme 
was used to generate events over a wide kinematic region.  The exchanged boson virtuality, $Q^2$, was required to be above 
1\,GeV$^2$, the squared photon--proton centre of mass energy, $W^2$, was required to be above 5\,GeV$^2$ and the fraction of the 
proton's momentum carried by the struck quark, $x$, was required to be greater than $10^{-7}$.  
The requirement on $x$ was due to technical issues running the generator at such low $x$ and will be relaxed in the future.  A test 
sample with an integrated luminosity of about 0.01\,pb$^{-1}$ was generated.  Investigation of the scattered electron showed that a 
kinematic peak at 3\,TeV is observed with the electron scattered at very low angles, essentially along the beamline axis.  This then 
requires detectors at shallow angles along the beam to measure the scattered electron.  The hadronic final state covered all polar angles 
and so a conventional colliding-beam detector will be needed.  However, at low $x$, the hadronic final state is produced at low angles 
and the angle is essentially zero and along the beamline for $x < 10^{-6}$.  This again requires instrumentation along the beamline.  A sketch of a 
detector system is shown in figure~\ref{fig:collidersketch}(b), where a series of spectrometers are needed in which the hadrons or electrons 
are bent away from the beamline and measured in dedicated detectors.

\begin{figure}
\begin{center}
\includegraphics[trim={0cm 0cm 1.5cm 7.2cm},clip,width=0.75\textwidth]{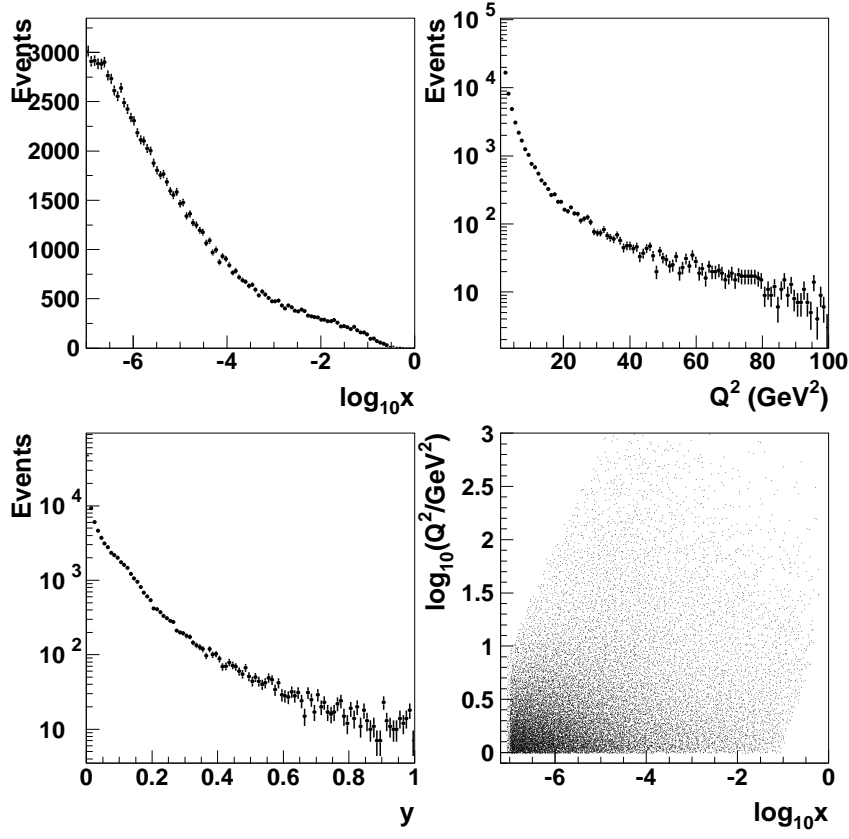}
\end{center}
\caption{
Kinematic variables, $x$, $Q^2$ and $y$, and the correlation between $Q^2$ and $x$ generated with the {\sc Aridane} Monte Carlo generator.  
For technical reasons, the events were limited to have an $x$ value above $10^{-7}$.  The number of events is plotted for a luminosity of 
about 0.01\,pb$^{-1}$.}
\label{fig:dis-kin}
\end{figure}

Distributions of the kinematic variables describing deep inelastic scattering, $x$, $Q^2$ and the inelasticity, $y$, are shown in 
figure~\ref{fig:dis-kin} along with the correlation of $Q^2$ and $x$.  As expected, the distributions peak at low values of $x$, $Q^2$ 
and $y$, with a strong correlation between $x$ and $Q^2$, also showing that once the technical Monte Carlo cut on $x$ is removed, 
values of $x \sim 10^{-8}$ for $Q^2 > 1$\,GeV$^2$ are possible.

\section{The photon--proton cross section at low $x$}
 
With such extended reach in $x$, VHEeP can be used to investigate the nature of the proton at very low parton momentum fractions, investigating 
the onset of saturation or other phenomena not described by standard QCD evolution of the parton densities.  A complementary approach is to 
consider the cross sections as a function of $1/x$ or the coherence length, $l$.  The coherence length is the distance over which a quark--antiquark 
pair can survive having come from a photon radiated by the incoming electron.  If the cross sections become the same as a function of $Q^2$, the 
photon states have had enough time to evolve into a universal size.  

\begin{figure}
\begin{center}
\includegraphics[trim={1.2cm 5.2cm 2.2cm 4.2cm},clip,width=0.75\textwidth]{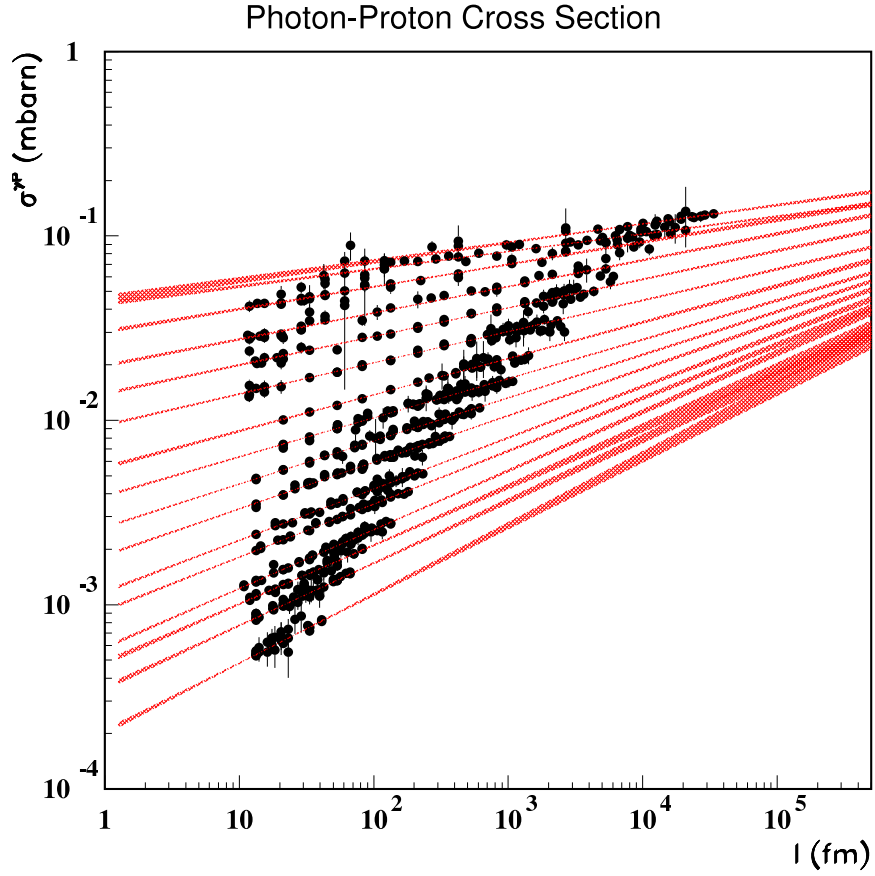}
\end{center}
\caption{
The photon--proton cross section, $\sigma^{\gamma P}$, shown versus the coherence length, $l$, for HERA data~\cite{1506.06042} at 
fixed values of $Q^2$ in the range 0.25 (top points) to 200\,GeV$^2$ (bottom points).  The data are fit for each $Q^2$ value with a 
function $\sigma_0 \, l^\lambda$.}
\label{fig:sigvsl}
\end{figure}

Initially the photon--proton cross section, $\sigma^{\gamma P}$, versus $l$, according to the definition by Stodolsky~\cite{stodolsky} 
($l = \hbar\,c/(2 \, \langle x \rangle \, M_P$)), is shown for the HERA data~\cite{1506.06042} in 
figure~\ref{fig:sigvsl}.  The data extend up to $l \sim 3 \times 10^4$\,fm which corresponds to $x \sim 3.5 \times 10^{-6}$.  The data are well 
described by a fit for each $Q^2$ value using the simple parametrisation $\sigma_0 \, l^\lambda$.  The figure also shows that the fits are converging 
for increasing $l$ and that it is expected that the cross section becomes independent of $Q^2$ at large $l$ (or small $x$).

\begin{figure}
\begin{center}
\includegraphics[trim={0.1cm 0.1cm 9.5cm 3cm},clip,width=0.75\textwidth]{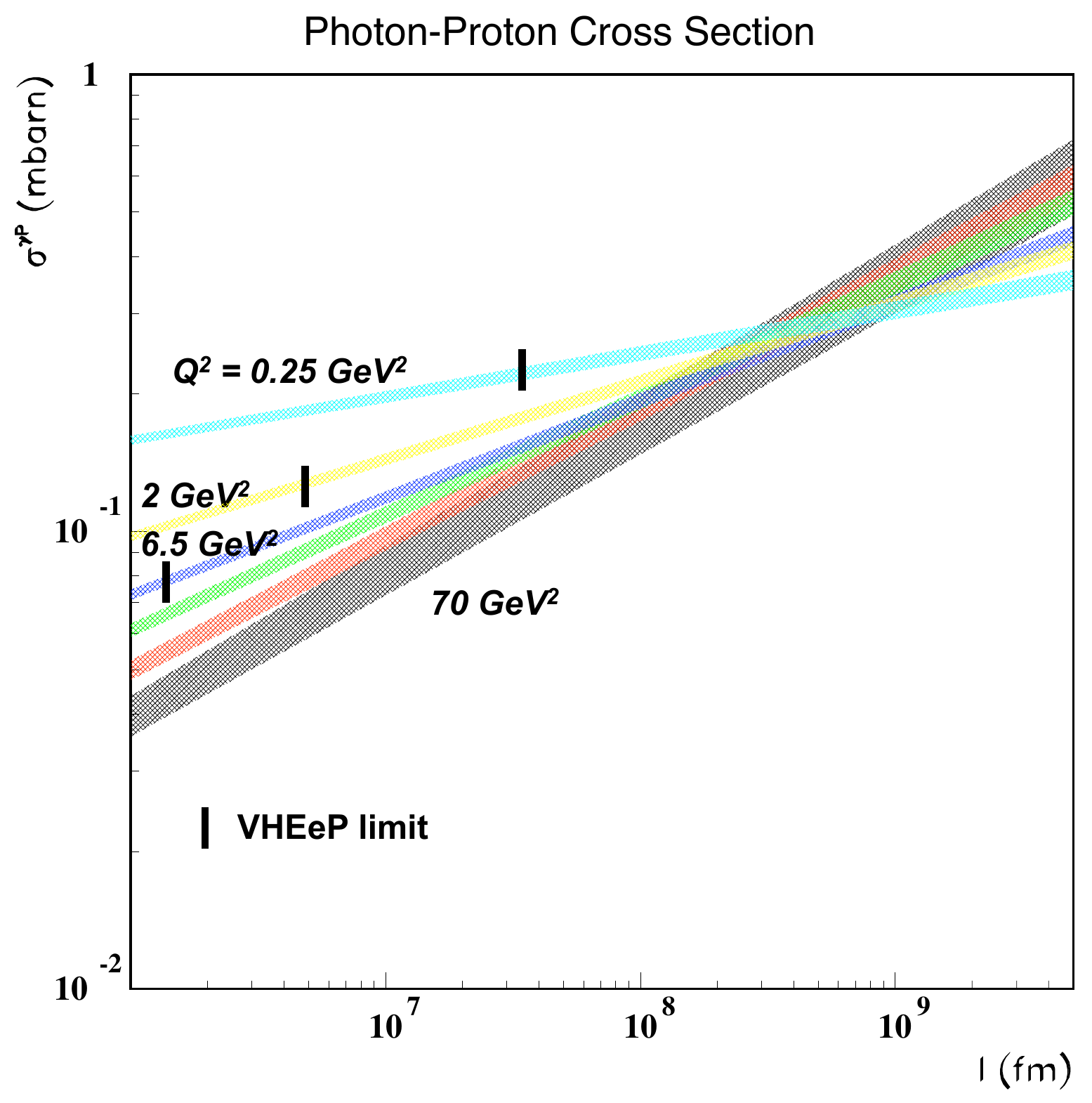}
\end{center}
\caption{
Fits of $\sigma_0 \, l^\lambda$ extrapolated with uncertainties to high $l$ (low $x$) for selected $Q^2$ values.
The upper kinematic limit of where VHEeP measurements will end is indicated (black vertical lines).
}
\label{fig:sigvsl-vheep}
\end{figure}

Figure~\ref{fig:sigvsl-vheep} shows the fits for selected $Q^2$ values extrapolated to higher $l$, where the fits cross at a value of 
$l \sim 3 \times 10^8$\,fm or $x \sim 3.5 \times 10^{-10}$.  An indication of the largest $l$ values that can be measured at VHEeP for 
each $Q$ value is also shown.  This demonstrates that measurements at $Q^2 < 1$\,GeV$^2$ will be very important in order to measure 
the region where the cross sections are expected to unify.  However, in the region $1 < Q^2 < 10$\,GeV$^2$, a very large number of events 
will be collected, even with the relatively low luminosities, and so able to constrain the fits for $l > 10^6$\,fm.  At higher $Q^2$, there will also 
be significant amounts of data, also at higher $l$ than was measured at HERA, and so will significantly constrain the extrapolation of the fits 
to higher $l$.  Further and more quantitative studies are needed, but already these initial investigations show that VHEeP will be able to constrain 
the onset of saturation in the proton.

\section{Summary and outlook}

In these proceedings, an idea for a very high energy electron--proton collider, VHEeP, has been presented.  Although of modest luminosities, 
1\,pb$^{-1}$/year, the centre-of-mass energy of 9\,TeV opens up a whole new kinematic region in $eP$ collisions.  Initial studies already indicate 
that effects such as saturation in the proton could be observed.  More work on developing a broad physics 
programme for VHEeP is ongoing.

\section*{Acknowledgements}

L L\"{o}nnblad is gratefully acknowledged for discussions using the {\sc Ariadne} Monte Carlo programme in this new kinematic 
regime.  E. Shaposhnikova is also gratefully acknowledged for discussions on the LHC beam parameters.

\end{document}